\documentstyle[prd,aps,psfig]{revtex}
\bibstyle{unsrt}
\begin{document}
 
\preprint{HD-THEP-97-59}  
\twocolumn[\hsize\textwidth\columnwidth\hsize\csname 
@twocolumnfalse\endcsname
\draft
 
\title{Time evolution of correlation functions in 
Non-equilibrium Field Theories}
\author{Lu\'{\i}s M. A.  Bettencourt$^{1,2}$ 
and Christof Wetterich$^1$}
\address{$^1$Institut f\"ur Theoretische Physik, University of
Heidelberg, Philosophenweg 16, 69120 Heidelberg, Germany}
\address{$^2$T6/T11, Theoretical Division, MS B288, 
Los Alamos National Laboratory, Los Alamos,New Mexico 87545} 
\date{\today}
\maketitle
\begin{abstract}
We investigate the non-equilibrium properties of an $N$-component scalar field 
theory. The time evolution of the correlation functions for an arbitrary 
ensemble of initial conditions is described by an exact functional 
differential equation. In leading order in the $1/N$ expansion
the system can be understood in terms of infinitely many conserved 
quantities. They forbid the approach to the canonical thermal distribution. 
Beyond leading order only energy conservation is apparent generically.
Nevertheless, we find a large manifold  of stationary distributions both 
for classical and quantum fields. They are the fixed points of the evolution 
equation. For small deviations of the correlation functions from
a large range of fixed points we observe stable oscillations. 
These results raise the question of if and in what sense the particular 
fixed point corresponding to thermal equilibrium dominates the large 
time behavior of the system.
\end{abstract}

\pacs{PACS Numbers : 11.15.Pg, 11.30.Qc, 05.70.Ln, 98.80.Cq 
\hfill   HD-THEP-97-59, LAUR - 974974} 

\vskip2pc]

The understanding of the time evolution for a quantum field theory with 
initial conditions away from thermal equilibrium is needed for a large 
variety of phenomena, ranging from inflationary cosmology in the 
early universe to laboratory experiments of defect formation.
First insights have been gained by perturbative methods and, in particular, 
by an investigation of the leading order in the $1/N$ expansion, with $N$ 
the number of components of the field \cite{MotRev,Mottola,InfBoya}.
This approximation has been applied to a variety of problems like 
decoherence of quantum states \cite{Mottola}, 
heating after cosmological inflation \cite{InfBoya}, 
disoriented chiral condensates in heavy ion collisions \cite{DCC}, 
decoherence and  photon pair creation from a homogeneous electric field 
\cite{photon}.
In most of this work a particle number basis was used in order to describe
the evolution of each field mode. From the mode evolution the two point 
function can in turn be constructed.

It is a severe limitation, however, that the leading $1/N$ approximation 
neglects scattering and cannot describe the approach of a system to a 
thermal equilibrium state. A systematic extension of the particle number 
formalism to finite $N$ seems a difficult task.
We pursue here an alternative strategy and investigate directly the time 
evolution of the correlation functions of the system. One starts, at $t=0$,
with an ensemble of initial values for the field variables and their time 
derivatives.
This specifies the initial correlation functions. At any later
time $t$ the correlation functions are, in principle, computable
from the microscopic field equations. In practice, we follow this time 
evolution by means of an exact functional differential evolution equation 
for the generating functional of the correlation functions \cite{Wetterich}.
Approximate solutions can be obtained by truncation.

Our main tool is the time dependent effective action
$\Gamma[\phi,\pi;t]$ \cite{Wetterich}. For field theories $\Gamma$ is a 
functional 
of the fields $\phi(x)$ and $\pi(x)= \partial_t \phi(x)$ and generates 
the one particle irreducible equal time correlation functions.
Its time evolution obeys
\begin{eqnarray}
\partial_t \Gamma[\phi, \pi] = - \left({\cal L_{ \rm cl}} 
+ {\cal L_{\rm q}} \right) \Gamma[\phi, \pi],
\label{e1}
\end{eqnarray}
where the first operator, ${\cal L_{\rm cl}}$, generates the 
classical evolution, whereas ${\cal L_{\rm q}}$ 
determines the dynamics of the quantum effects.   
We will deal with the specific example of 
a $O(N)$ symmetric field theory for real scalars 
$\chi_a(x)$, $a=1,...,N$, obeying the microscopic equations of motion of a  
$\phi^4$-model 
\begin{eqnarray}
\partial_t^2 \chi_a = \partial_i \partial_i \chi_a - m^2 \chi_a - 
{\lambda\over 2} \chi_b \chi_b \chi_a. \label{e2}
\end{eqnarray}  
The arguments of $\Gamma$ are the mean values of the field 
$\phi_a = \langle \chi_a \rangle$ and its time derivative $\pi_a = 
\langle \dot \chi_a \rangle $ (in the presence of sources).
From \cite{Wetterich,QWett}, one infers
\begin{eqnarray}
&& {\cal L}_{\rm cl} = \int d^D x \left\{ \pi_a(x) 
{\delta \over \delta \phi_a(x)} + \phi_a(x)\left(\nabla^2 - m^2 \right)
{\delta \over \delta \pi_a(x)} \nonumber \right. \\ 
&&- \left. {\lambda \over 2} 
\left[\phi_b(x) \phi_b(x) \phi_a(x) +  \phi_a(x) 
G^{\phi \phi}_{bb}(x,x) \nonumber \right. \right.
\\ && + \left. \left.  2 \phi_b(x) G^{\phi \phi}_{ba}(x,x) 
\nonumber  \right. \right. \\ && -  \left. \left. 
\int d^Dx_1  d^Dx_2  d^Dx_3 G^{\phi \psi}_{ai}(x,x_1)
G^{\phi \psi}_{bj}(x,x_2) \nonumber \right. \right. \\
&& \times \left. \left.
G^{\phi \psi}_{bk}(x,x_3) {\delta^3 \Gamma \over 
\delta \psi_i(x_1) \delta \psi_j(x_2) \delta \psi_k(x_3)} \right ] 
{\delta \over \delta \pi_a(x) } \right\} ,  \\
&& {\cal L}_{\rm q} 
= {\lambda \over 8} \hbar^2\int d^D x~ \phi_a(x)
{\delta \Gamma \over \delta \pi_b(x)} 
{\delta \Gamma \over \delta \pi_b(x)} 
{\delta  \over \delta \pi_a(x)},
\label{e3}
\end{eqnarray}
with $\psi \in \{\phi, \pi \}$ and repeated index summations implied.
The exact field dependent propagator $G$, interpreted as a matrix with 
multi-indices $(\psi,i,x)$, is the inverse of the second functional 
derivative of $\Gamma$,
\begin{eqnarray}
\left( G^{-1} \right)^{\psi \psi'}_{ij} (x,y) = { \delta^2 \Gamma 
\over \delta \psi_i(x) \delta \psi_j' (y)}.
\end{eqnarray}

We emphasize that the minimum of $\Gamma[\phi,\pi;t]$ describes
the time evolution of  a cosmological scalar field like the inflaton. The true
dynamical equation (\ref{e1}) for non-equilibrium situations differs,
in general, from the variation of some equilibrium effective
action or the classical action. The latter are at best an approximation to 
Eq.~(\ref{e1}) in certain limiting situations not too far from equilibrium. 
An understanding of the structure of the dynamical equations 
following from (\ref{e1}) is of crucial importance for an understanding of 
cosmological models like inflation.

We assume an $O(N)$-symmetric initial distribution for which all terms 
with odd powers of the fields in $\Gamma$ vanish. This property 
remains conserved by the evolution. Similarly, we consider a translational 
and rotational invariant situation.
The most general form of $\Gamma$ consistent with these 
symmetries can be expanded in terms of the Fourier components of the fields
\begin{eqnarray}
\Gamma &=& {1 \over 2} \int {d^D q \over (2 \pi)^D } 
\Bigl\{  A(q)\phi_a^*(q)\phi_a(q)  +   B(q) \pi_a^*(q) \pi_a(q) 
\nonumber \Bigr. \\ && \Bigl. +  2 C(q) \pi_a^*(q)
\phi_a (q) \Bigr\} 
+ {1\over 8} \int {d^D q_1 \over (2 \pi)^D} {d^D q_2 \over (2 \pi)^D} 
{d^D q_3 \over (2 \pi)^D}  \nonumber \\ 
&& \Bigl\{ 
u(q_1,q_2,q_3)  \phi_a(q_1) \phi_a(q_2) \phi_b(q_3) 
\phi_b(-q_1-q_2-q_3) \Bigr. \nonumber \\
&& \Bigl. 
+ v(q_1,q_2,q_3) \pi_a(q_1) \phi_a(q_2) \phi_b(q_3) 
\phi_b(-q_1-q_2-q_3) \Bigr. \nonumber \\
&& \Bigl. 
+w(q_1,q_2,q_3)\pi_a(q_1) \pi_a(q_2) \phi_b(q_3) 
\phi_b(-q_1-q_2-q_3) \Bigr. \nonumber \\
&& \Bigl. + s(q_1,q_2,q_3) \left[\pi_a(q_1) \pi_b(q_2) \phi_a(q_3) 
\phi_b(-q_1-q_2-q_3) \right. \Bigr. \nonumber \\
&& \Bigl. \left. \qquad - \pi_a(q_1) \pi_a(q_2) \phi_b(q_3) 
\phi_b(-q_1-q_2-q_3) \right]  \label{e4} \Bigr.  \\
&& \Bigl. + y(q_1,q_2,q_3) \pi_a(q_1) \pi_a(q_2) \pi_b(q_3) 
\phi_b(-q_1-q_2-q_3) \Bigr. \nonumber \\
&& \Bigl. +  z(q_1,q_2,q_3) 
\pi_a(q_1) \pi_a(q_2) \pi_b(q_3)  \pi_b(-q_1-q_2-q_3) \Bigr\} 
\nonumber \\
&& + \ldots \nonumber
\end{eqnarray}
Here $A,B$ and $C$ are functions of $q^2$ and the quartic couplings $u,v$, 
 {\it etc.}, 
depend on the invariants that can be constructed from the three independent
momenta $q_1,q_2,q_3$. 

The  evolution equation can also be expanded in powers
of the fields and the exact equation for the inverse propagator reads
\begin{eqnarray}
&&\partial_t A(q) = 2 \omega_q^2 C(q), \nonumber \\
&&\partial_t B(q) = - 2 C(q) - 2 \gamma_q  B(q), \label{e5}   \\
&&\partial_t C(q) =   \omega_q^2 B(q) -  A(q) - \gamma_q C(q). \nonumber 
\end{eqnarray}
Here, the time dependent frequency, $\omega_q$, and the coefficient 
$\gamma_q$ are:
\begin{eqnarray}
&& \omega_q^2 = q^2 +m^2 + {N+2 \over 2}  \lambda
\int  {d^D p \over (2 \pi)^D} G(p) \label{e6} \\ 
&& \quad - {N+2 \over 8} \lambda \int {d^D p \over (2 \pi )^D}
{d^D p' \over (2 \pi )^D} G(p) G(p') G(p+p'+q)  \nonumber  \\ 
&& \Bigl[ 4 u(p,p',-p-p'-q) - 3 v(p,p',-p-p'-q) c(p)  \nonumber \Bigr. \\  
&& \Bigl. \qquad \qquad + 2 w(p,p',-p-p'-q) c(p) c(p') \Bigr. \nonumber \\ 
&& \Bigl. \qquad \qquad - y(p,p',-p-p'-q) c(p) c(p') c(p+p'+q) 
\Bigr], \nonumber  \\
&& \gamma_q = { N+2 \over 8} \lambda \int {d^D p \over (2 \pi )^D}
{d^D p' \over (2 \pi )^D} G(p) G(p') G(p+p'+q)  \nonumber  \\ 
&& \Bigl[  v (q,p,p') - 2 w(q,p,p') c(p) +3 y(q,p,p') c(p) c(p') 
\Bigr. \nonumber \\ && \Bigl. \qquad \qquad 
-4 z (q,p,p') c(p) c(p') c(p+p'+q)   \Bigr],
\label{e7}
\end{eqnarray} 
with 
\begin{eqnarray}
c(q) = { C(q) \over B(q) }
\end{eqnarray}	
and $G(q)$ the exact $\phi-\phi$ propagator for vanishing fields 
\begin{eqnarray}
G(q) &=& {B(q) \over A(q)B(q) -C^2(q)} \equiv  {B(q) \over \alpha^2(q)}.
\label{e8}
\end{eqnarray}  
We suppose that all momentum integrals are ultraviolet regularized by a 
cutoff $\Lambda$, ($q^2 < \Lambda^2$).  For the simple system of $N$ 
coupled anharmonic oscillators ($D=0$) the momentum labels and integrals 
should be omitted. From (\ref{e5}) one obtains the flow equations 
for $G(q)$, $c(q)$ and $\alpha^2(q)$, 
\begin{eqnarray}
&& \partial_t \alpha^2(q) = -2 \gamma_q \alpha^2(q), \nonumber\\
&& \partial_t G(q) = -2 c(q) G(q),  \label{e9}\\
&& \partial_t c(q) = \omega^2_q - {1 \over \alpha^2(q) G^2(q)} 
+ \gamma_q c(q) + c^2(q). \nonumber
\end{eqnarray} 

For large $N$, with $\lambda, u,v,w,s,y,z$ scaling like $1/N$,
one may expand in $1/N$ as a small parameter. We observe that 
$\gamma_q \sim 1/N$ and $\omega_q$ becomes independent of 
$u,v,\ldots,$ in leading order. In the limit $N \rightarrow \infty$
the system of evolution equations for the inverse two-point functions
$A,B$ and $C$ is closed.

For a solution of Eq.~(\ref{e5}) at finite $N$, however, one needs the time
dependent four point functions. Their evolution equations involve, 
in turn the six point functions. The exact system
is not closed and for any practical purpose we have to proceed to some 
approximation. In the present note we simply omit all contributions from 
1 PI six point vertices. Furthermore for $D \geq 1$, we neglect the 
momentum dependence of the quartic couplings. With this truncation
the evolution equations for the quartic couplings are given by
\begin{eqnarray}
&&\partial_t u =   {\bar \omega}^2 v - \lambda \hbar^2 {\bar B}^3 {\bar c}^3 
\nonumber \\
&& + 4 \lambda {\bar B} {\bar c} \left[ 1 -  
u {\cal S}_0 + {v \over 2} {\cal S}_1 - { (N+2) w - (N-1)s 
\over 2 (N+8)} {\cal S}_2   
\right],  \nonumber \\
&& \partial_t v =   2 {\bar \omega}^2  w  
- 4 u - {\bar \gamma} v - 3 \lambda \hbar^2 {\bar B}^3 {\bar c}^2 
\nonumber \\ 
&& + 4 \lambda {\bar B} \left[ 1 -  u {\cal S}_0  + {v \over 2} {\cal S}_1 
- { (N+2) w -  (N-1)s \over 2 (N+8)} {\cal S}_2 \right] \nonumber  
\\ &&  \qquad - 2 \lambda {\bar B} {\bar c} \left[ v {\cal S}_0 
- {12 w + 2(N-1) s 
\over N+8} {\cal S}_1 + y {\cal S}_2 \right], \label{e10} \\
&&\partial_t w =  3 {\bar \omega}^2 y - 3v - 2 {\bar \gamma} w 
 - 3 \lambda \hbar^2 {\bar B}^3  {\bar c} \nonumber \\
&& \qquad - 2 \lambda {\bar B}  
\left[  v {\cal S}_0 - {12 w + 2(N-1) s \over N+8} {\cal S}_1  
+ y {\cal S}_2 \right] \nonumber \\
&& - 2 \lambda {\bar B} {\bar c}\left[  { (N+2) w - 
 (N-1) s \over N+8}{\cal S}_0  - y {\cal S}_1 + 2 z {\cal S}_2 \right], 
\nonumber \\
&&\partial_t s =  2 {\bar \omega}^2 y - 2 v  - 2 {\bar \gamma} s
- 2 \lambda  \hbar^2 {\bar B}^3 {\bar c} \nonumber \\
&& -\lambda {2 {\bar B} \over N+8} \Bigl[ (N + 6) v {\cal S}_0 
- 2 (4 w +N s) {\cal S}_1 + (N+6) y {\cal S}_2 \Bigr] \nonumber \\
&& - {4 \lambda {\bar B}{\bar c} \over N+8}\Bigl[ s {\cal S}_0 -2y {\cal S}_1 
+4 z {\cal S}_2 \Bigr], \nonumber \\ 
&& \partial_t y =  4 {\bar \omega}^2 z - 2w - 3 {\bar \gamma} y  
- \lambda \hbar^2 {\bar B}^3\nonumber \\
&& -  2 \lambda {\bar B}\left[{(N+2) w -  (N-1) s \over N+8}  
{\cal S}_0  - y {\cal S}_1 + 2 z {\cal S}_2 \right], \nonumber \\
&&\partial_t z =  - y  - 4 {\bar \gamma}  z, \nonumber 
\end{eqnarray}
with
\begin{eqnarray}
{\cal S}_{\rm k} = {N+8 \over 2}\int {d^D p \over (2 \pi )^D}   
G^2(p) c^{\rm k} (p).
\end{eqnarray}
For $D \geq 1$ there remains at this stage an ambiguity
at which momentum the couplings $u,v,$ etc. and therefore the
barred quantities $ {\bar \omega}^2,{\bar \gamma}, {\bar c}, \bar B$,  
should be evaluated.(For $D=0$ the bars should be omitted.) 
One may choose some appropriate momentum average to be discussed
below.
From the truncated evolution 
equation of the quartic couplings  we have omitted some terms 
$\sim 1/N^2$ which involve two-loop integrals. 
The neglected six point functions can also be treated 
consistently as being $\sim 1/N^2$.
On the other hand, the difference 
$u(q,p,p') - u(0,0,0)$ appears  in the contributions $\sim 1/N$. 
Nevertheless, the effects of this momentum dependence are partly averaged 
out by the momentum integrations.
Thus our set of evolution equations~(\ref{e10}) contains for 
$D=0$ all contributions in the next to leading order $1/N$. 
This extends to arbitrary $D$ only if the momentum dependence of the quartic 
couplings can be neglected. 
  
The microscopic field equations~(\ref{e2}) conserve the total energy $E$ for 
every choice of initial state 
\begin{eqnarray}
&& E = \int d^D x \left[ {1 \over 2} \partial_t \chi_a  \partial_t \chi_a 
+{1 \over 2} \partial_i \chi_a  \partial_i \chi_a  
+ {1 \over 2} m^2 \chi_a \chi_a  \nonumber \right. \\ 
&& \qquad \qquad \qquad \left. + {1\over 8} \lambda 
\chi_a \chi_a \chi_b \chi_b \right].
\end{eqnarray}
Therefore the average energy per volume
\begin{eqnarray}
&& \epsilon = {\langle E \rangle  \over \Omega } = \epsilon_1 + \epsilon_2, 
\nonumber \\
&& \epsilon_1 = {N \over 2} \int {d^D q \over (2 \pi)^D} \Biggl\{
B^{-1}(q) + \Bigl[ q^2 + m^2 + c^2(q)  \Bigr. \Biggr. \nonumber \\ 
&& \qquad \qquad \Biggl. \Bigl.
+ {N+2 \over 4} \lambda 
\int  {d^D p \over (2 \pi)^D} G(p) \Bigr] G(q) \Biggr\},  \\
&& \epsilon_2 = - {N \left( N+2 \right)\over 8 } \lambda  
\int {d^D q_1 \over (2 \pi)^D} {d^D q_2 \over (2 \pi)^D}  
{d^D q_3 \over (2 \pi)^D} G(q_1) G(q_2) \nonumber \\ 
&& \times G(q_3) G(-q_1- q_2 -q_3)
\Bigl[ u(q_1,q_2,q_3) - v(q_1,q_2,q_3) c(q_1)  \nonumber \Bigr. \\ 
&& \Bigl. + w(q_1,q_2,q_3) c(q_1)c(q_2)     
- y(q_1,q_2,q_3)c(q_1)c(q_2) c(q_3) \nonumber \Bigr. \\
&& \Bigl. + z(q_1,q_2,q_3)c(q_1)c(q_2)c(q_3)c(-q_1 -q_2 -q_3) \Bigr], 
\nonumber
\end{eqnarray} 
must be conserved by the exact evolution equations.
We have verified that the truncated evolution equations 
(\ref{e5}), (\ref{e10}) conserve
$\epsilon$ for $D=0$ and arbitrary $N$. For $D \geq 1$ this holds 
automatically in leading order in the $1/N$ expansion. For finite 
$N$ our truncation conserves energy 
only provided the averages ${\bar \omega}^2$, $\bar B$, $\bar c$, 
$\bar \gamma$, are chosen appropriately such that they obey the 
relation
\begin{eqnarray}
&& \partial_t \epsilon = -{N (N+2) \over 8} \lambda 
\int {d^D q \over (2 \pi)^D} {d^D p \over (2 \pi)^D}  
{d^D p' \over (2 \pi)^D} G(q) G(p) \nonumber  \\ 
&& \times G(p') G(p + p' +q)  \Biggl\{ \left[ 
\left( {\bar \omega}^2 - \omega_q^2 \right) + \left({\bar \gamma} 
- \gamma_q \right) c(q)\right] \Biggr.  \nonumber \\
&& \Biggl. \left[ v -2w c(p) + 3 y c(p) c(p') 
-4 z c(p) c(p') c(p+p' +q) \right]  \Biggr. \nonumber \\
&& \Biggl. 
+ 4 \lambda {\bar B} \left({\bar c} - c(q) \right) \Bigl[
1 - u {\cal S}_0 + {v \over 2}
\left( {\cal S}_1 + c(p) {\cal S}_2 \right) \Bigr. 
\Biggr. \nonumber \\
&& \Biggl. \Bigr.
- { \left( N+2 \right) w -  \left(N-1 \right) s \over 
2 \left(N+8 \right)} \left[ {\cal S}_2 + c(p) c(p') {\cal S}_0 \right]
\Bigl. \Biggr.  \nonumber  \\ 
&& \Biggl. \Bigl. 
- {6 w +  \left(N-1 \right) s \over \left( N+8 \right) } c(p) {\cal S}_1
+ {y \over 2} \left[ c(p)  {\cal S}_2   
+ c(p) c(p') {\cal S}_1 \right] \Bigr. \Biggr. \nonumber \\
&& \Biggl. \Bigl.
- z {\cal S}_2 - {1 \over 4} \hbar^2 {\bar B}^2 \left( {\bar c } - c(p) 
\right) \left( {\bar c} - c(p') \right) \Bigr] \Biggr\} = 0. 
\end{eqnarray}
For practical purposes it may be sufficient that this relation holds
in a time averaged sense. Whereas $\epsilon_1$ is separately conserved 
in the leading $1/N$ approximation the flow of energy between $\epsilon_1$ 
and $\epsilon_2$ can be attributed to scattering.
  
The fixed points of the system Eqs.~(\ref{e5}), (\ref{e10}) correspond
to stationary probability distributions. They 
obey, for $\omega_q^2> 0$,
\begin{eqnarray}
&& A_*(q) = \omega_q^2 B_*(q), \qquad   C_*(q) = 0,   \nonumber \\
&& v_*=0, \qquad  y_*=0,  \nonumber \\ 
&& u_* =    {\lambda {\bar B}_* 
+  \left ( {\bar \omega}^2/2 \right) w_* \over 1 
+ \lambda {\bar B}_*{\cal S}_0}, \label{e11} \\
&& w_* =   {2 {\bar \omega}^2 z_* +  {N-1 \over N+8} \lambda {\bar B}_* 
{\cal S}_0 s_* - {\lambda \over 2}\hbar^2  {\bar B}_*^3  
\over  1+ {N+2 \over N+8} \lambda {\bar B}_* {\cal S}_0}. \nonumber
\end{eqnarray}
We observe a large manifold of fixed points since Eqs.~(\ref{e11}) have 
solutions for arbitrary $B_*(q)$, $s_*$ and $z_*$ !
This property seems not to be an artifact of the truncation. 
It persists for momentum dependent four point functions. 
The  fixed point manifold 
becomes even larger if one includes the six point functions and seems 
to characterize any higher (finite) polynomial truncation.
Classical thermodynamic equilibrium, at temperature $T$ 
(for $\hbar=0$) corresponds to the 
particular point in this manifold:
\begin{eqnarray}
&& B_{\rm eq}(q) = \beta =1/T, \quad  C_{\rm eq}(q)= 0, \nonumber \\ 
&& A_{\rm eq}(q)= \beta (\omega_q^2)_{\rm eq} = 
G_{\rm eq}^{-1} (q),  \label{e12}\\
&& v_{\rm eq} = w_{\rm eq} = s_{\rm eq} = y_{\rm eq} = z_{\rm eq} = 0, 
\nonumber \\
&& u_{\rm eq} = \lambda \beta \left( 1 - {\cal S}_0 u_{\rm eq} \right).
\nonumber
\end{eqnarray}
Here the fixed point conditions for $A_{\rm eq}$ and $u_{\rm eq}$ constitute 
classical Schwinger-Dyson equations (corresponding to a 
Euclidean $D$-dimensional quantum field theory at zero temperature) for 
the effective propagator (mass and kinetic terms) 
and the effective quartic coupling, respectively. 
In quantum field theory ($\hbar=1$) the thermal 
equilibrium fixed point does not correspond to 
$w_{\rm eq} = s_{\rm eq} = z_{\rm eq} = 0$, anymore. The Schwinger-Dyson 
equations have now to be evaluated in a $D+1$ dimensional quantum field 
theory with periodic boundary conditions in 'Euclidean time'.
The difference between the quantum and classical fixed points is due to 
the nonzero Matsubara frequencies. We observe, nevertheless, a rather similar 
structure of the classical and quantum dynamical systems. 
In particular, we find in the quantum system a fixed point, which differs 
from the classical equilibrium fixed point only by 
$z_* = \lambda \hbar^2 {\bar B}_*^3/(4 {\bar \omega} ^2)$, 
instead of $z_{\rm eq}=0$. 
For this fixed point, all correlation functions involving only 
$\phi$ are the same as for classical equilibrium. Inversely, for every 
fixed point of the quantum system there is a corresponding one for the 
classical system, differing only by a shift in $z_*$. Up to this shift, the 
quantum equilibrium could be reached by the classical system !    

The central issue of this work is the question of whether and 
how  the fixed points are approached for $t \rightarrow \infty$. 
We start with the leading behavior for  $N\rightarrow \infty$, where 
\begin{eqnarray} 
\gamma_q = 0, \quad \omega_q^2 = q^2 + m^2 + \Delta m^2,
\end{eqnarray}
with 
\begin{eqnarray}
\Delta m^2 = {\lambda N \over 2}\int {d^D p \over (2 \pi)^D} G(p) 
\end{eqnarray}
and
\begin{eqnarray}
\partial_t \omega^2_q =  
\partial_t \Delta m^2 
= - \lambda N \int  {d^D p \over (2 \pi )^D} 
~{C(p) \over \alpha^2 (p)} \equiv \beta_\omega.
\label{e13}
\end{eqnarray}  
In this limit the Fourier components of the determinant $\alpha^2(q)$ 
become an infinite set of conserved quantities !
(cf. Eqs.~(\ref{e11}), (\ref{e13}).) 
Obviously, the system can reach thermal equilibrium only for a very 
particular set of initial conditions, namely if $\alpha^2(q) = 
G^{-1}_{\rm eq}(q)$, for every $q$. 

\begin{figure}
\centerline{\psfig{file=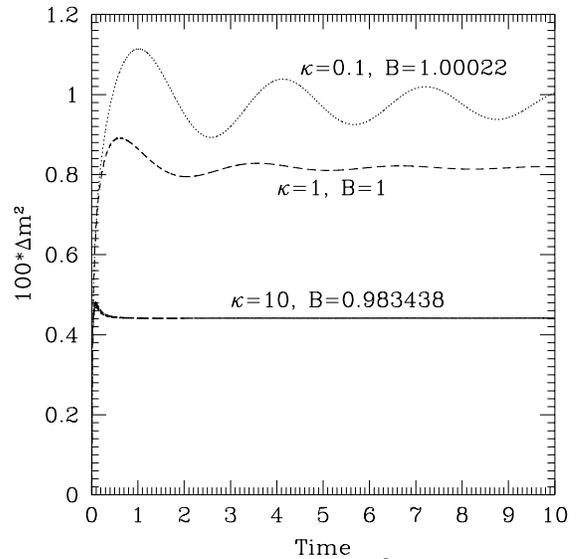,width=3.0in}}
\caption{The time evolution of $ \Delta m^2 $ for initial conditions
with $ G(q)=1/(q^2 + \kappa^2 m^2)$ and $\alpha^2(q)= B(q)/G(q)$,
$m^2 =1$ and $\lambda N=2$.  $B(q)=B$ is 
chosen so that the energy $\epsilon_1$ is the 
same in all three cases.  The ultraviolet 
cutoff was chosen to be $\Lambda=100$ and the total number of modes
is $2000$.
For long times $\Delta m^2$ tends to different constant values for initial 
conditions with the same $\epsilon_1$ but different $G(q)$ and 
$\alpha^2(q)$.} 
\label{fig1}
\end{figure}

Nevertheless, for large times and $D\geq 1$,  $\Delta m^2$ 
(or $\omega^2_q$) tends to a constant value as can be seen in 
Fig.~\ref{fig1}, for $D=3$.
This phenomenon, called 'dephasing' in \cite{Mottola}, 
is due to the superposition of rapidly  oscillatory functions 
$C(q)$ on the right hand side of Eq.~(\ref{e13}).
In fact, for $C(q) = \psi_R(q) \cos(2 \omega_q t) + 
\psi_I(q) \sin(2 \omega_q t)$,  
$\alpha^2(q) = \omega^2_q/\zeta^2(q)$ and using $y=\omega_q t = 
t \sqrt{q^2 + m_R^2}$ as integration variable ($m_R^2 = m^2  
+ \Delta m^2$) one has, for $D=3$,
   
\begin{eqnarray}
&& \beta_\omega(t) = - {\lambda N \over 2 \pi^2 t}
\int_{m_R t}^\infty  dy~ \zeta^2(y)~
\sqrt{1 - {m_R^2 t^2 \over y^2}}~ \nonumber \\
&& \qquad \qquad \times \left(\psi_R(y) 
\cos(2 y) + \psi_I(y) \sin(2 y) \right).
\label{e14}
\end{eqnarray}
For large $m_R t$ the amplitudes $\psi_R$ and $\psi_I$ vary only slowly 
with $y$ since the leading oscillations have been factorized out.
This also should hold for $\zeta^2(y)$. Typically these quantities are smooth 
functions of $y/m_R t$ such that their $y$-derivatives 
are suppressed by $1/m_R t$. ( For $q^2 \rightarrow \infty$ we will assume 
that $\psi_R(q), \psi_I(q)$ and $\zeta^2(q)$ remain finite.) The oscillatory 
behavior of the integrand prevents the $y$-integral from increasing like 
$m_R t$ and $\beta_\omega$ therefore vanishes for $t \rightarrow \infty$. 
We emphasize that in general 
the asymptotic constant value $\Delta m^2$ depends on the initial 
conditions and is {\em not equal}
to the temperature dependent mass renormalization in thermal equilibrium.

For a constant $\Delta m^2$ the dynamical system simplifies considerably 
and we  obtain  a free oscillator for each momentum  $q$, 
with frequency $\omega_q$.
The amplitude of the  oscillations determines in turn the distance 
to the fixed point $C_* =0$ and $A_* -(\omega_q^2)_* B_* =0$.
Thus for  generic initial conditions the 
$N \rightarrow \infty$ system does not  reach asymptotically the fixed 
point manifold Eqs.~(\ref{e11}). If there is a 'dynamical fixed point' for
$\Delta m^2$ the correlation functions of the system end up oscillating.
In addition to $\alpha(q)$ the system with constant $\Delta m^2$ has another 
infinite  set of conserved quantities, namely $A(q) + \omega_q^2 B(q)$. 
The late time behavior of the system is 
clearly not ergodic. Ergodicity is usually thought to be 
a sufficient, albeit not necessary, condition
for subsystems of a large closed system to approach the canonical 
distribution \cite{Parisi}.
 
The generic oscillatory behavior   
is also apparent from the Hamiltonian structure which governs the 
dynamical evolution of the two point functions 
of the $N \rightarrow \infty$ system.
In terms of canonical coordinates $Q(q)$ and $P(q)$, 
\begin{eqnarray}
A(q) &=& \alpha^2(q) P^2(q) + 1/Q^2(q), \quad B(q)= \alpha^2(q) Q^2(q), 
\nonumber \\
C(q) &=&  - \alpha^2 (q) P(q) Q(q), \qquad G(q) = Q^2(q) 
\end{eqnarray}
the equations of motion become
\begin{eqnarray}
&& \partial_t Q(q) = P(q), \nonumber \\
&& \partial_t P(q) = - \omega^2_q Q(q) + {1 \over \alpha^2(q)  Q^3(q)}, \\
&& \omega_q^2 = q^2 +m^2 + {\lambda N \over 2} \int 
{d^D p \over (2 \pi)^D}Q^2(p). 
\label{e15}
\end{eqnarray}
The effective Hamiltonian
\begin{eqnarray}
&& H_{\rm eff} = \epsilon_1  =  
N \int {d^D q \over (2 \pi )^D} \left\{{ P^2(q) \over 2} + V_q[Q] 
\right\}, \label{e16} \\
&& V_q[Q] = {1 \over 2} (q^2 +m^2) Q^2(q) + 
{1\over 2} {1 \over  \alpha^2 (q) Q^2(q)} \nonumber  \\
&& +  \lambda  {N  \over 8 } 
\int {d^D p \over (2 \pi )^D} Q^2(p)Q^2(q), \nonumber
\end{eqnarray} 
is conserved in time\footnote{This effective Hamiltonian has been found 
in a different approach in \cite{Mottola}. We emphasize that in leading 
order in the $1/N$ expansion there is no difference between classical 
and quantum statistical systems and $H_{\rm eff}$ appears in both of them.}. 
The potential is bounded from below and the minimum 
corresponds to the fixed point solution Eq.~(\ref{e11}), {\it i.e.} 
$P(q) = 0$, 
$\omega_q^2 Q^4(q) = 1/\alpha^2(q)$. For any initial value of $H_{\rm eff}$ 
larger than the minimum,  the system cannot approach the fixed point and 
must remain oscillatory.  

Let us next turn to the system for finite $N$ and investigate the behavior 
in the vicinity of the fixed points. For this purpose we identify $u,v,$ 
{\it etc.},
with the couplings at zero momentum and therefore ${\bar \omega}^2 = 
\omega_0^2$ , ${\bar B} = B(0)$, and so on. 
In the following we concentrate on classical statistics ($\hbar=0$).     
We  consider 
the linear equations for small deviations from one of the fixed points 
(\ref{e11}), with $\delta G(q) = G(q) - G_*(q)$, {\it etc.}. For simplicity we 
choose $w_* = s_*=0$, whereas $B_*(q)$ remains arbitrary 
such that   the 
fixed point does not necessarily coincide with thermal equilibrium (\ref{e12}).
In terms of directions orthogonal to the fixed point manifold

\begin{eqnarray}
&& \hat a(q) = \delta \alpha^2(q) + { 2 \over \omega_q^2 G^3(q) }  \delta G(q)
+ { 1 \over  \omega_q^4 G^2(q)} \delta \omega^2_q, \nonumber \\
&& \hat u = \delta u - {1 \over 2} { \omega_0^4 G(0) \over \omega_0^2 G(0) 
+ \lambda { \cal S}_0} w  \label{e17} \\
&& \quad + {\lambda \over \left( \omega_0^2 G(0) + 
\lambda { \cal S}_0 \right)^2 }  \left( \omega_0^2 \delta G(0) + G(0) 
\delta \omega_0^2 + 
\lambda \delta {\cal S}_0 \right), \nonumber \\
&& \hat z = z - {1 \over 2 \omega_0^2} \left[ \left( { 6 \over N+8} + 
{N+2 \over N+8} {\lambda \over u}  {1 \over \omega_0^2 G(0) }\right) w  
\right. \nonumber \\ 
&& \qquad \left. + { (N-1) \over N+8} \left( 1 - {\lambda \over u } 
{1 \over \omega_0^2 G(0) } \right) s \right], \nonumber 
\end{eqnarray} 
the linear dynamical equations read  
\begin{eqnarray}
&& \partial_t \delta G(q) = - 2 G(q) c(q), \nonumber \\
&& \partial_t \hat a(q) = - {\lambda {\cal T}_0(q) \over 2 \omega_q^2 G^2(q)} v
- {4 \over \omega_q^2 G^2(q)} c(q) + {1 \over \omega_q^4 G^2(q)} \partial_t 
\omega_q^2,  \nonumber  \\
&& \partial_t c(q) = \omega_q^4 G^2 (q) \hat a (q), \nonumber  \\
&& \partial_t \hat u =  {2 + 4 \sigma \over 1+ \sigma}  u 
c(0) +  {5 + 4 \sigma \over 2 (1 + \sigma) } \omega_0^2 v
\nonumber  \\
&& \qquad -  {3 \over 2(1 + \sigma)} \omega_0^4 y 
+ {1 \over 1 + \sigma} { u \over \omega_0^2 }
\partial_t \omega_0^2 + u^2 \partial_t S_0, \nonumber \\
&& \partial_t v = - 4 \left( 1 +\sigma \right) \hat u 
+ 4 u \omega_0^2 G^2(0) \hat a(0), \label{e18}  \\
&& \partial_t w = 3 \omega_0^2 y -  \left(3 + 2 \sigma \right) v,\nonumber  \\
&& \partial_t s = 2 \omega_0^2 y - 2 \left( 1 + { N+6 \over N+8} 
\sigma \right) v, \nonumber  \\
&& \partial_t y = 4 \omega_0^2 \hat z, \nonumber \\ 
&& \partial_t \hat z = - {\left( 5 + \sigma \right) \over 2} y 
+ { 3 \over 2 \omega_0^2} \left[ 1  
+ \sigma + { 10 N + 44 \over 3 (N+8)^2} \sigma^2 \right] v,
\nonumber
\end{eqnarray}
with
\begin{eqnarray}
&& \sigma = {\lambda {\cal S}_0 \over \omega_0^2 G(0) } = 
{\lambda \over u \omega_0^2 G(0)} -1, \nonumber \\
&& {\cal T}_0(q) = { N+2 \over 2}  \int {d^D p \over (2 \pi )^D}
{d^D p' \over (2 \pi )^D} G(p)  \\ 
&& \qquad \qquad \qquad \qquad \times G(p') G(p+p'+q). \nonumber  
\end{eqnarray}
All coefficients multiplying the linear excitations should be evaluated 
at the fixed point. The excitations $\delta G(q)$, $w$ and $s$ are considered 
coordinates on the fixed point manifold (\ref{e11}) and do not appear 
on the right-hand side of Eq.~(\ref{e18}). ( Note that $\partial_t \omega_q^2$ 
and $\partial_t S_0$ do not involve  these quantities.)
We can therefore discuss the subsystem $\left\{ \hat a(q), c(q), 
\hat u, v, y, \hat z \right\}$ separately. We first consider a scenario, for 
$D \geq 1$, 
where  $\omega_0^2$ and $S_0$ are almost independent of time due to 
the averaging out of 
fluctuations with different momenta. In the limit $\partial_t 
\omega_0^2=0$, 
$\partial_t S_0 = 0$, the system for $\chi = \left\{ \hat a(0), c(0), 
\hat u, v, y, \hat z \right\}$ becomes closed, $\partial_t \chi_i 
= A_{ij} \chi_j$. 
On the other hand, the anharmonic oscillator for $D=0$ obeys
\begin{eqnarray}
&& \partial_t \omega_0^2 = -\kappa_\omega \lambda (N+2) G(0) c(0) 
\left( 1- {6 \over N+8} {\sigma \over 1+\sigma } \right), \nonumber \\
&& \partial_t  {\cal S}_0 = -2 \kappa_{\cal S} (N+8) G^2(0) c(0),
\end{eqnarray} 
with $\kappa_\omega =\kappa_{\cal S} \equiv 1 $. In order to interpolate 
between both situations we keep $\kappa_\omega $ and $\kappa_{\cal S}$ 
as free parameters.
The matrix $A_{ij}$ determines the behavior near the 
fixed point. 

For the free system, with $\lambda = 0$, 
$\sigma= 0$, the six eigenvalues 
$\xi$ of 
$A$ are $\pm 2 i \omega_0, \pm 2 i \omega_0,\pm 4 i \omega_0$, as 
expected for the harmonic oscillations around the fixed point.
For the  interacting system  the eigenvalues are  
determined by a cubic equation, $ \eta^3 + b \eta^2 + c \eta + d = 0$,
in $\eta=\xi^2/4 \omega_0^2$, where
\begin{eqnarray}
&& b = 6 \left\{ 1 + {5 \sigma \over 12} + {\tau \over 3} + \Sigma_1 
 \right\} , \nonumber \\
&& c = 9 \left\{ 1 + {13 \over 18} \sigma  + {1 \over 18} 
\left[ 2 - {3 \left( 5 N 
+ 22 \right) \over (N+8)^2} \right] \sigma^2  \right. \nonumber \\  
&& \left. \qquad \qquad  + {1 \over 9}
\left( 4 - \sigma \right) \tau  + {5 (2 + \sigma) \over 3} 
\Sigma_1 + {1\over 9} \Sigma_2 \right\},  \\
&& d= 4 \left\{\left[ 1 + \sigma  + {1\over 8} \left[ 2 - {3 \left( 5 N 
+ 22 \right) \over (N+8)^2} \right] \sigma^2 \right] \left( 1 + 6 \Sigma_1 
\right)  \right. \nonumber \\ 
&& \left. \qquad \qquad - {1 \over 8} \left( 5 
+ 11 \sigma + 2 \sigma^2 \right) \tau  + {5 + \sigma \over 8} \Sigma_2 
\right\}, \nonumber
\end{eqnarray}
with 
\begin{eqnarray}
&& \tau= \lambda u {\cal T}_0 \left( 4 \omega_0^2 \right)^{-1}, 
\nonumber  \\
&& \Sigma_1 = {\kappa_\omega \rho \sigma \over 12} {N+2 \over N+8}
\left( 1 - {6 \over N+8} { \sigma \over 1+ \sigma } \right), \nonumber \\
&& \Sigma_2 = \rho \sigma \tau  \Biggl[ 2  
\kappa_{\cal S} \Biggr.  \\
&& \Biggl. \qquad \qquad + \kappa_\omega 
\left( {N+2  \over N+8} \right)  \left( 1 - 
{6 \over N +8} {\sigma \over 1 + \sigma} \right) \Biggr], \nonumber \\
&& \rho = {N + 8 \over 2} { 1 + \sigma \over \sigma} u G^2(0). \nonumber
\end{eqnarray}
The behavior of the linearized system depends on the sign of the 
discriminant, $K = {1 \over 27} b^2 c^2 - {4 \over 27} c^3 - d^2 
- {4 \over 27} b^3 d + {2 \over 3} b c d$. For $K \geq 0$, the three roots 
$\eta_i$ are all real and negative, for positive $b,c$ and $d$. 
The linearized system is oscillatory with eigenvalues 
$\xi_{i\pm} = \pm 2 i \omega_0 \sqrt{\vert \eta_i \vert } $. 
For negative $K$ two 
roots $\eta$ are complex and conjugate to each other. Since the real
part of $\xi$ does not vanish in this case
one concludes that two eigenvalues $\xi$ must have a positive real part. 
They dominate the evolution for $t \rightarrow \infty$.
The fixed point is unstable for $K < 0$.

The fact that the eigenvalues of $A_{ij}$ occur in pairs with opposite sign 
is a direct consequence of time reflection symmetry of the microscopic 
equations of motion. The correlation functions are eigenstates of time 
reflection
$\hat T$- in our case $c,v$ and $y$ have odd $\hat T$-parity.
One infers that the linear system obeys $\hat T A \hat T = -A$ with 
$\hat T = {\rm diag}(+1,-1,+1,-1,-1,+1)$. This implies that for any eigenvalue
$\xi$ of $A$ there also exists an eigenvalue $-\xi$. One concludes that an 
approach to the equilibrium fixed point cannot be seen in linear order.    
For time reflection invariant systems the only way an equilibrium fixed 
point could be approached is a purely oscillatory behavior in linear order 
with vanishing real parts of all eigenvalues of the 'stability matrix' 
$A$. Beyond linear order it is then conceivable that the oscillations are 
damped by the non-linear flow - at least for a large class of trajectories. 
Of course, $\hat T$-invariance implies that for any trajectory with flow 
towards the fixed point there must also exist a trajectory with flow away 
from it.  Nevertheless, for a nonlinear flow it seems possible that only
a very special class of excitations around the fixed point leads to 
unstable behavior. The relative volume of the space of unstable 
excitations ( as compared to the stable excitations) could shrink to zero 
as the number of degrees of freedom goes to infinity. 

We should point out that the general oscillatory behavior found in our 
linear analysis is not in contradiction with the damping rates computed 
in linear response theory for classical or quantum field theories in 
thermodynamic equilibrium \cite{AB}.
Indeed, a linear superposition of a continuum of frequencies with
suitable amplitudes can easily lead
to a damped behavior, as illustrated by the identity
$\int_{- \infty}^{+ \infty} {d \omega \over \pi } { \gamma \over 
\omega^2 + \gamma^2} e^{ i \omega t} = e^{-\gamma \vert t \vert}$.
The damping of $\Delta m^2$ shown in Fig.~\ref{fig1} also constitutes a 
good example. It is therefore conceivable that
on a microscopic level the oscillations are never damped and the 
system never reaches the thermodynamic equilibrium fixed point in 
a strict sense. Nevertheless, macroscopic quantities may well approach 
equilibrium values for $t\rightarrow \infty$.

Let us first consider $D=0$ where $\kappa_\omega = 
\kappa_{\cal S} = \rho = 1, \tau = {N+2 \over 2 (N+8)^2} 
{\sigma^2 \over 1 +\sigma}$, such that the cubic equation 
only depends on one parameter, $\sigma$. In an expansion in 
powers of $1/N$ we find in lowest order $K=0$,
with roots  $\eta_i = - (1 + { \sigma \over 2}),
- (1 + { \sigma \over 2}),  - 4 (1 + { \sigma \over 2})$.
This is very similar to the oscillations of the free system. However, 
the relevant frequency is shifted to 
\begin{eqnarray}
\omega \sqrt{ 1 + \sigma/2} =  m^2 + {3\over 4}N \lambda G,
\end{eqnarray}
with the
temperature dependent 'propagator' ($T= B_*^{-1}$)
\begin{eqnarray}
G= 2 T \left( \sqrt{ m^4 + 2 N \lambda T } + m^2 \right)^{-1}.
\end{eqnarray}
In order to discuss the stability of the equilibrium fixed point we 
have to proceed to next to leading order $1/N$ where
\begin{eqnarray} 
K = {\sigma^2  \over N} \left(1  + { \sigma \over 2} \right)^3 
\left( {47 + 23 \sigma - 13 \sigma^2 \over 1 + \sigma } \right)  
+O\left({1 \over N^2} \right).  
\end{eqnarray}
This is  positive for $\sigma < \sigma_c \approx 2.982$. Using the leading 
order relation
\begin{eqnarray}
&& \sigma = \left( \sqrt{ 1 + \nu^2} + \nu \right)^{-2}, \quad 
\nu = {m^2 \over \sqrt{ 2 N \lambda T}},
\end{eqnarray}
one has $\sigma <1$ for positive $m^2$ and no unstable mode is present. 
On the other hand, for negative $m^2$ there is a critical value 
$\nu_c \approx - 4.39$, beyond which $\sigma$ exceeds $\sigma_c$.
This leads to the most interesting conclusion that for negative $m^2$ 
and $\lambda N < 0.026 {m^4 \over T}$, the system is repelled from
thermodynamic equilibrium asymptotically, despite the fact that 
the number of coupled degrees of freedom can be arbitrarily large 
($N\rightarrow \infty$) ! For a given initial average energy per mode 
($\sim T$) the stability of thermodynamic 
equilibrium requires a minimal coupling strength !

In contrast, for $\kappa_\omega = \kappa_{\cal S} = 0$ ($D > 0$), 
we find already in lowest order in an $1/N$ expansion the roots 
$ \eta = -(1+ \sigma/2), -1, 
(-4 + 2 \sigma)$, with 
\begin{eqnarray}
K= {3 \over 4}\sigma^2 \left( 1 + {\sigma \over 2} \right)^2 \left( 
1+ { 2 \over 3} \sigma \right)^2 > 0.
\end{eqnarray}
The linear system is purely oscillatory.
Since $K$ is a continuous function of $1/N$ and $\tau$ (note that $\tau 
\sim 1/N$ ) this remains true for small $1/N$. 
For example in next to leading order we obtain
\begin{eqnarray}
&& \Delta K_{1/N} =  {\tau \over 4} 
\left( 1 + {2 \over 3} \sigma \right) \left (6 \sigma ^4 + 38 
\sigma ^3 + 111 \sigma^2  \right.  \\
&& \left. + 150 \sigma + 72 \right)   
- {5 \over 24 N}  \sigma^3 \left( 1 + {2 \over 3} \sigma \right)
\left( \sigma^2 + 24 \sigma + 36 \right). \nonumber  
\end{eqnarray}
This conclusion also extends to non-vanishing $\kappa_\omega$ and 
$\kappa_{\cal S}$, as long as $\kappa_\omega \sim 1/N$. 
We conclude that the equilibrium fixed
point has no unstable direction for linear fluctuations.    
By simple continuity considerations this stability can be extended
to a whole region in the fixed point manifold around the 
thermal equilibrium point.

The discussion of the nonlinear behavior for large 'distances' from the 
fixed point manifold  remains outside the scope of this note.
We only report here briefly some results for $D=0$ where Eqs.~(\ref{e5}),
(\ref{e13}) reduce to nine coupled ordinary differential equations, which
we have solved numerically. The stable small oscillations around 
the fixed points near classical thermal equilibrium can easily be observed, 
being determined by three relevant frequencies.  In this case we
have not observed any nonlinear damping. As the amplitudes
of the oscillations increase the stable behavior continues for a while.
The time averaged mean value corresponds now to
a 'dynamical fixed point', which is close to but not part of the 
fixed point manifold. For a Gaussian initial distribution not too
far from the fixed point manifold ( and for $m^2 > 0$) this behavior
is approached after a certain time. Finally, for large distances from 
the fixed point manifold the behavior becomes much more complicated. 
Typically a very large range of frequencies becomes important after a certain 
time interval and the problem becomes stiff. Analogous behavior has been 
observed in other anharmonic oscillator systems treated in the 
$1/N$ expansion \cite{0D}. We will address these interesting 
phenomena elsewhere.

In conclusion, the step beyond the leading order $1/N$ expansion seems crucial
for the understanding of non-equilibrium field theories. One finds in 
leading order infinitely many unphysical conserved quantities which 
prevent the system from approaching thermal equilibrium. They do not
survive the inclusion of scattering, beyond leading order.   New mysteries 
have appeared,
however. This concerns, in particular, the existence of a continuous 
manifold of stationary distributions within which the thermal 
equilibrium distribution is just a particular point. It has also 
become clear that small deviations from the fixed points are not 
damped in the context of linear fluctuation analysis. The understanding 
of the thermalization of
isolated time reversible systems described by field theories 
remains an unsolved challenge.   
As long as such basic questions are not fully understood the use 
of equilibrium field equations in cosmology should be 
regarded with  caution.    

This work was partially supported by the {\it Deutsche 
Forschungsgemeinshaft} and the European Science Fundation.


\begin{thebibliography}{99} 

\bibitem{MotRev} F. Cooper, S. Habib, Y. Kluger, E. Mottola
Phys. Rev. D {\bf 55}, 6471 (1997).

\bibitem{Mottola} S. Habib, Y. Kluger, E. Mottola and J. P. Paz
Phys. Rev. Lett.  {\bf 76}, 4460 (1996).

\bibitem{InfBoya} D. Boyanovsky, D. Cormier, H. J. de Vega and R. Holman
Phys. Rev. D {\bf 55}, 3373 (1997).

\bibitem{DCC} F. Cooper, Y. Kluger, E. Mottola and J. P. Paz,
Phys. Rev. D {\bf 51} 2377 (1995); D. Boyanovsky, H. J. de Vega
and R. Holman, Phys.Rev. D {\bf 51} 734 (1995).

\bibitem{photon} F. Cooper {\it et al.}, Phys. Rev. D {\bf 50} 2848 
(1994).

\bibitem{Wetterich} C. Wetterich, Phys. Rev. Lett. {\bf 78} 3598
(1997).

\bibitem{QWett} C. Wetterich, Phys. Rev. E {\bf 56} 2687 (1997);
 H. Nachbagauer, hep-th/9703018.

\bibitem{Parisi} See {\it eg.} G. Parisi, {\it Statistical Field Theory}, 
Addison-Wesley Publishing  Company, New York, 1988. 

\bibitem{AB} G. Aarts and J. Smith, Phys. Lett. B {\bf 393} 395 (1997);
W. Buchm\"uller and A. Jakov\'ac,  Phys. Lett. B {\bf 407} 39 (1997)

\bibitem{0D} B. Mihaila, J. F Dawson and F. Cooper, Phys. Rev. D {\bf 56}
5400 (1997).

\end{thebibliography}
\end{document}